\documentstyle[preprint,eqsecnum,tighten,amssymb,aps]{revtex}

\begin{document}

\title{Two-Photon Algebra Eigenstates: 
A Unified Approach to Squeezing}
\author{{\large\sc C. Brif} \thanks{Email: 
phr65bc@phys1.technion.ac.il}}
\address{Department of Physics, Technion -- Israel Institute of 
Technology, Haifa 32000, Israel}
   \maketitle
\vspace*{0.3cm}
\centerline{to appear in {\sc Annals of Physics}}
\vspace*{-0.7cm}
	\begin{abstract}
We use the concept of the algebra eigenstates that provides a 
unified description of the generalized coherent states (belonging 
to different sets) and of the intelligent states associated with a
dynamical symmetry group.  
The formalism is applied to the two-photon algebra and the 
corresponding algebra eigenstates are studied by using the 
Fock-Bargmann analytic representation. This formalism yields a
unified analytic approach to various types of single-mode 
photon states generated by squeezing and displacing transformations. 
	\end{abstract}

\section{Introduction}

Coherent states (CS) associated with various dynamical symmetry 
groups are important in many problems of quantum physics 
\cite{KlSk,Per86,Gil:rev}. Actually, there are three distinct ways 
in which CS for a Lie group can be defined \cite{Gil:rev}.

In the general group-theoretic approach developed by 
Perelomov \cite{Per} and Gilmore \cite{Gil}, the CS are generated
by the action of group elements on a reference state of a group
representation Hilbert space. These states (called the generalized 
CS) have a number of remarkable properties that make them very 
useful in description of many quantum phenomena 
\cite{KlSk,Per86,Gil:rev}. The most important features of the 
coherent-state systems are their overcompleteness and their 
invariance under the action of group representation operators. The 
last property means that the CS transform among themselves during 
the evolution governed by Hamiltonians for which the corresponding 
Lie group is the dynamical symmetry group.

The second approach deals with states defined as eigenstates of a
lowering group generator. Attention was mainly paid to  
eigenstates of the lowering generator $K_{-}$ for different 
realizations of SU(1,1) \cite{BG,EOCS,Agar88,Buz90,BBA:qo,I:qso}.

The third way in which CS can be defined is associated with the
optimization of uncertainty relations for Hermitian generators of a 
group \cite{Schro,Arag,RBA,NiTr,BHY_YH,PrAg,Trif,Puri,BBA:jpa,GeGr}.
States that minimize uncertainty relations are called 
intelligent states (IS) or minimum-uncertainty states. Ordinary
IS \cite{Arag} provide an equality in the Heisenberg uncertainty 
relation while generalized IS \cite{Trif,Puri} do so in the 
Robertson uncertainty relation \cite{Robert}. The IS are determined 
by some type of the eigenvalue equation \cite{Jackiw,Trif,Puri}, and 
the lowering-generator eigenstates are in fact a particular case of 
the IS, corresponding to equal uncertainties of two Hermitian 
generators. 

In the special case of the Heisenberg-Weyl group $H_{3}$
\cite{Weyl} whose generators are the boson annihilation and creation
operators $a$ and $a^{\dagger}$ and the identity operator $I$, the
first and second definitions coincide. The Glauber CS 
$|\alpha\rangle$ \cite{Gla} can be defined as eigenstates of the
lowering generator, $a|\alpha\rangle = \alpha|\alpha\rangle$, and 
also as states generated by the displacement operator $D(\alpha)$
(representing group elements) acting on the vacuum state $|0\rangle$,
	\begin{equation}
|\alpha\rangle = D(\alpha) |0\rangle = \exp(\alpha a^{\dagger} -
\alpha^{\ast} a) |0\rangle .  \label{1.1}
	\end{equation}
At the same time, the Glauber CS $|\alpha\rangle$ are the IS for the
field quadratures $X_{1}=(a^{\dagger}+a)/2$ and 
$X_{2}=i(a^{\dagger}-a)/2$, i.e., they minimize the Heisenberg 
uncertainty relation $\Delta X_{1} \Delta X_{2} \geq 1/4$. The 
uncertainties are equal, $\Delta X_{1} = \Delta X_{2} = 1/2$, when
the expectation values are calculated for the $|\alpha\rangle$
states. In this sense, the CS $|\alpha\rangle$ are a special case
of the canonical squeezed states \cite{SS}. For the squeezed states,
the fluctuations in one quadrature are reduced on account of growing 
fluctuations in the other (conjugate) quadrature. The canonical 
squeezed states can be considered as the generalized IS for the 
Heisenberg-Weyl group \cite{NiTr,Trif}.

For more complicated groups, e.g., for SU(1,1), the different
definitions lead to distinct states. The Perelomov CS for the 
SU(1,1) Lie group, obtained by the action of the group elements on 
the reference state \cite{Per86}, 
and the Barut-Girardello states, defined as the eigenstates of 
the SU(1,1) lowering generator $K_{-}$ \cite{BG}, are quite 
different. However, the concept of squeezing can be naturally
extended to the SU(1,1) group, and the squeezing properties of the
SU(1,1) ordinary and generalized IS have been widely discussed 
\cite{WodEb,Ger85_88,Hil87_89,Agar88,Buz90,%
NiTr,BHY_YH,PrAg,Trif,Puri,BBA:jpa,GeGr}.

In Perelomov's definition, different sets of the CS are obtained
for different choices of the reference state. The usually used sets
of the CS (the standard sets, as we refer to them) correspond to the
cases when an extreme state of the representation Hilbert space
(e.g., the vacuum state of the quantized field mode) is chosen as 
the reference state \cite{Gil:rev}. In general, this choice of the 
reference state leads to the sets consisting of states with 
properties closest to those of classical states \cite{Per86}. On the
other hand, the IS show a variety of nonclassical properties, such 
as squeezing and sub-Poissonian photon statistics. In the case of 
the SU(1,1) Lie group, the standard set of Perelomov's CS and the
set of the ordinary IS have an intersection \cite{WodEb,BBA:jpa}.
Both these types of states form subsets of the generalized IS 
\cite{Trif}.

In this paper we develop a formalism that provides a unified
description of different types of coherent and intelligent states.
We introduce the concept of algebra eigenstates (AES) which are 
defined for an arbitrary Lie group as eigenstates of elements of 
the corresponding complex Lie algebra. We show that different sets 
of the generalized CS (both standard and nonstandard) can be 
equivalently defined as the AES. Moreover, the ordinary and
generalized IS for Hermitian generators of a Lie group form a subset
of the AES associated with this group. On the basis of the 
algebra-eigenstate formalism, we use analytic methods that enable
us to treat different types of states (including the standard and 
nonstandard CS and the IS) in a unified way. This unified 
description is also applicable for investigating more complicated 
states obtained by the action of unitary group transformations on  
the IS. Such states can be considered as (nonstandard) generalized 
CS with the reference state being an intelligent state. 

In the present work we apply the general formalism to the 
two-photon group $H_{6}$ \cite{Gil:rev} that enables us
to obtain the unified description of single-mode photon states
generated by displacing and squeezing transformations. We use 
the Fock-Bargmann analytic representation \cite{FB} based on the 
standard set of the Glauber CS. In this analytic representation 
the eigenvalue equation that determines the two-photon AES 
becomes a linear homogeneous differential equation.
Then the powerful theory of analytic functions is applied for
studying various types of photon states and relations between them.

In Sec.\ 2 we develop the group-theoretic formalism of the AES
for an arbitrary Lie group. The Fock-Bargmann representation of the
two-photon AES is derived in Sec.\ 3. By using this 
representation, we find entire analytic functions representing 
different types of photon states. In Sec.\ 4 we consider  
displaced and squeezed Fock states. The superpositions of the 
Glauber CS (the Schr\"{o}dinger-cat states) and their squeezed and 
displaced versions are discussed in Sec.\ 5. The two-photon IS for
the SU(1,1) subgroup of $H_{6}$ are considered in Sec.\ 6. 
We introduce the states which are generated by squeezing and 
displacement of the IS. We also touch on the question of the 
production of various two-photon AES.

\section{The general theory of the algebra eigenstates}

Let $G$ be an arbitrary Lie group and $T$ its unitary irreducible 
representation acting on the Hilbert space ${\cal H}$. By choosing 
a fixed normalized reference state $|\Psi_{0}\rangle \in {\cal H}$, 
one can define the system of states $\{ |\Psi_{g}\rangle \}$,
	\begin{equation}
|\Psi_{g}\rangle = T(g) |\Psi_{0}\rangle , \mbox{\hspace{0.8cm}}
g \in G, \label{2.1}
	\end{equation}
which is called the coherent-state system. 

The isotropy (or maximum-stability) subgroup $H \subset G$ consists 
of all the group elements $h$ that leave the reference state 
invariant up to a phase factor,
	\begin{equation}
T(h) |\Psi_{0}\rangle = e^{i\phi(h)} |\Psi_{0}\rangle , 
\mbox{\hspace{0.8cm}} | e^{i\phi(h)} | =1 ,
\mbox{\hspace{0.3cm}} h \in H .		\label{2.2}
	\end{equation}
For every element $g \in G$, there is a unique decomposition of $g$ 
into a product of two group elements, one in $H$ and the other in 
the quotient (or coset) space $G/H$,
	\begin{equation}
g = \Omega h , \mbox{\hspace{0.8cm}} g \in G, \;\; h \in H, \;\; 
\Omega \in G/H .	\label{2.3}	
	\end{equation}
It is clear that group elements $g$ and $g'$ with different $h$ and 
$h'$ but with the same $\Omega$ produce the coherent states which 
differ only by a phase factor: $|\Psi_{g}\rangle = e^{i\delta} 
|\Psi_{g'}\rangle$, where $\delta =\phi(h) -\phi(h')$. Therefore a 
coherent state $|\Psi_{\Omega}\rangle$ is determined by a point 
$\Omega = \Omega(g)$ in the quotient space $G/H$. 

One can see from this group-theoretic procedure for the construction
of the generalized CS that the choice of the reference state 
$|\Psi_{0}\rangle$ firmly determines the structure of the 
coherent-state set. An important class of coherent-state sets 
corresponds to the quotient spaces $G/H$ which are homogeneous
K\"{a}hlerian manifolds. Then $G/H$ can be considered as the phase 
space of a classical dynamical system, and the mapping $\Omega 
\rightarrow |\Psi_{\Omega}\rangle \langle\Psi_{\Omega}|$ is the 
quantization for this system \cite{Ber}. It means that the 
quantization is performed via the CS \cite{Per86}.

Let us consider the Lie algebra $\frak{G}$ of the group $G$ (here 
and in the what follows we will call algebra the complex extension 
of the real algebra, i.e., the set of all linear combinations of 
elements of the real algebra with complex coefficients). 
The isotropy subalgebra ${\frak B}$ is defined as the set of 
elements $\{ {\frak b} \}$, ${\frak b} \in {\frak G}$, such that
	\begin{equation}
{\frak b} |\Psi_{0}\rangle = \lambda |\Psi_{0}\rangle . \label{2.4}
	\end{equation}
Here $\lambda$ is a complex eigenvalue. If the isotropy subgroup $H$
is nontrivial, then the isotropy subalgebra ${\frak B}$ will be
nontrivial too. By acting with $T(g)$ on both sides of Eq.\ 
(\ref{2.4}), we obtain
	\begin{equation}
T(g) {\frak b} T^{-1}(g) T(g)|\Psi_{0}\rangle =
\lambda T(g)|\Psi_{0}\rangle . \label{2.5}
 	\end{equation}
This leads to the eigenvalue equation
	\begin{equation}
{\frak g} |\Psi_{g}\rangle = \lambda |\Psi_{g}\rangle , \label{2.6}
	\end{equation}
where $|\Psi_{g}\rangle = T(g) |\Psi_{0}\rangle$ is a coherent 
state, and the operator ${\frak g} = T(g) {\frak b} T^{-1}(g)$ is 
an element of the algebra $\frak{G}$.
We see that the generalized CS are the eigenstates of the elements 
of the complex algebra.

Now, let us choose a basis 
$\{ {\frak K}_{1},{\frak K}_{2},\ldots,{\frak K}_{p} \}$ for a 
$p$-dimensional Lie algebra $\frak{G}$. Then an element of the 
complex algebra can be written as the Euclidean scalar product
in the $p$-dimensional vector space,
	\begin{equation}
{\frak g} = \bbox{\beta} \cdot\bbox{{\frak K}} 
= \beta_{1}{\frak K}_{1} + \beta_{2}{\frak K}_{2} + \cdots +
\beta_{p}{\frak K}_{p} ,    \label{2.7}
	\end{equation}
where $\beta_{1},\beta_{2},\ldots,\beta_{p}$ are arbitrary complex 
coefficients. Then the AES are defined by the eigenvalue equation:
	\begin{equation}
\bbox{\beta}\cdot\bbox{{\frak K}} 
|\Psi(\lambda,\bbox{\beta})\rangle = 
\lambda |\Psi(\lambda,\bbox{\beta})\rangle .   \label{2.8}
	\end{equation}
The comparison of Eqs.\ (\ref{2.6}) and (\ref{2.8}) shows that
the generalized CS can be defined as the AES, and a specific set of 
the CS is obtained for the appropriate choice of the parameters 
$\beta$'s. More precisely, let a state $|\Psi(\lambda,\bbox{\beta})
\rangle$ belong to a specific set of the CS corresponding to the 
reference state $|\Psi_{0}\rangle$ that satisfies Eq.\ (\ref{2.4}). 
Then the parameters $\beta$'s must satisfy the condition 
$\bbox{\beta}\cdot\bbox{{\frak K}} = T(g) {\frak b} T^{-1}(g)$, 
$\forall g \in G$.
Note that the definition (\ref{2.8}) of the AES does not depend 
explicitly on the choice of the reference state $|\Psi_{0}\rangle$. 
Hence it is possible to treat the CS defined as the AES in a quite 
general way, regardless of the set to which they belong. 

An important property of the generalized CS is the identity 
resolution:
	\begin{equation}
\int d\mu(\Omega) |\Psi_{\Omega}\rangle \langle\Psi_{\Omega}| = I ,   
\label{2.9}
	\end{equation}
where $I$ is the identity operator in the Hilbert space ${\cal H}$, 
and $d\mu(\Omega)$ is the invariant measure in the homogeneous 
quotient space $G/H$. Then any state $|\Psi\rangle \in {\cal H}$ can 
be expanded in the coherent-state basis $|\Psi_{\Omega}\rangle$,
	\begin{equation}
|\Psi\rangle = \int d\mu(\Omega) f(\Omega) |\Psi_{\Omega}\rangle ,   
\label{2.10}
	\end{equation}
where $f(\Omega) = \langle\Psi_{\Omega}|\Psi\rangle$, and
	\begin{equation}
\langle\Psi|\Psi\rangle = \int d\mu(\Omega) |f(\Omega)|^{2} .  
\label{2.11}
	\end{equation}
If we restrict the consideration to the square-integrable Hilbert 
space then the integral in (\ref{2.11}) must be convergent. Since 
the CS are not orthogonal to each other, the CS themselves can be 
expanded in their own basis. 

Now, let us represent all the AES in the standard coherent-state 
basis. In what follows we will consider 
only the simplest cases in which the quotient space $G/H$ 
corresponding to the standard set is a homogeneous K\"{a}hlerian 
manifold that can be parametrized by a single complex number 
$z$, so we write the standard generalized CS $|\Psi_{\Omega}\rangle$ 
in the form $|z\rangle$. Then Eq.\ (\ref{2.10}) reads
	\begin{equation}
|\Psi(\lambda,\bbox{\beta})\rangle = \int d\mu(z) 
f(\lambda,\bbox{\beta};z^{\ast}) |z\rangle  .  \label{2.12}  	
	\end{equation}
The function $f(\lambda,\bbox{\beta};z) = \langle z^{\ast}|
\Psi(\lambda,\bbox{\beta})\rangle$ can be decomposed into two 
factors: $f(\lambda,\bbox{\beta};z) = 
{\cal N}(z) \Lambda(\lambda,\bbox{\beta};z)$. 
Here ${\cal N}(z)$ is a normalization factor such that 
$\Lambda(\lambda,\bbox{\beta};z)$ is an entire analytic function 
of $z$ defined on the whole complex plane or on part of it. Such 
analytic representations are well studied \cite{FB,Per86} for 
the standard coherent-state bases of the simplest Lie groups. In 
these simplest cases the elements of the Lie algebra act in the 
Hilbert space of entire analytic functions as linear
differential operators. Then the eigenvalue equation (\ref{2.8}) 
is converted into a linear homogeneous differential equation. By 
solving this equation, we obtain the entire analytic functions 
$\Lambda(\lambda,\bbox{\beta};z)$ representing the AES 
$|\Psi(\lambda,\bbox{\beta})\rangle$ in the standard coherent-state 
basis $|z\rangle$. 

The standard set of the CS is a particular case of the wide 
system of the AES. Other particular cases of the AES are the sets 
of the ordinary and generalized IS. Any two quantum observables 
(Hermitian operators in the Hilbert space) $A$ and $B$ obey the 
Robertson uncertainty relation \cite{Robert}
	\begin{equation}
(\Delta A)^{2} (\Delta B)^{2} \geq \frac{1}{4} 
( \langle C \rangle^{2} +  4 \sigma_{AB}^{2} ) , 
\mbox{\hspace{0.7cm}} C=-i[A,B] ,
\label{2.13}
	\end{equation}
where the variance of $A$ is $(\Delta A)^{2} = \langle A^{2} \rangle 
- \langle A \rangle^{2}$, $(\Delta B)^{2}$ is defined similarly, 
the covariance of $A$ and $B$ is $\sigma_{AB} = \frac{1}{2} 
\langle AB+BA \rangle - \langle A \rangle\langle B \rangle$, and the 
expectation values are taken over an arbitrary state in the Hilbert 
space. When the covariance of $A$ and $B$ vanishes, 
$\sigma_{AB} = 0$, the Robertson uncertainty relation reduces to the 
Heisenberg uncertainty relation,
	\begin{equation}
(\Delta A)^{2} (\Delta B)^{2} \geq \frac{1}{4}    
\langle C \rangle^{2} .  \label{2.14}
	\end{equation}
The states which provide an equality in the Heisenberg uncertainty
relation (\ref{2.14}) are called the ordinary IS \cite{Arag} and 
the states which minimize the Robertson uncertainty relation 
(\ref{2.13}) are called the generalized IS \cite{Trif}. It is clear 
that the ordinary IS form a subset of the generalized IS. 
The generalized IS for operators $A$ and $B$ are determined from 
the eigenvalue equation \cite{Trif,Puri}
	\begin{equation}
(\eta A + i B) |\lambda,\eta\rangle 
= \lambda |\lambda,\eta\rangle ,  \label{2.15}
	\end{equation}
where the parameter $\eta$ is an arbitrary complex number, and 
$\lambda$ is a complex eigenvalue. For the particular case of real 
$\eta$, the eigenvalue equation (\ref{2.15}) determines the 
ordinary IS for operators $A$ and $B$. Then the equation can be 
written in the form \cite{Jackiw}
	\begin{equation}
(A + i\gamma B) |\lambda,\gamma\rangle = 
\lambda |\lambda,\gamma\rangle ,   \label{2.16}
	\end{equation}
where $\gamma$ is a real parameter. By comparing Eqs.\ (\ref{2.15}) 
and (\ref{2.16}) with Eq.\ (\ref{2.8}), we see that the IS 
for any two Hermitian group generators form a subset of the AES of 
the group.

The generalized IS for the quadratures $X_{1}$ and $X_{2}$ coincide  
with the canonical squeezed states \cite{Trif}. The concept of  
squeezing is naturally related also to the IS associated with
the SU(2) and SU(1,1) Lie groups \cite{WodEb,Ger85_88,Hil87_89,%
Agar88,Buz90,NiTr,BHY_YH,PrAg,Trif,Puri,BBA:jpa,GeGr}. At the last 
years there 
is a great interest in the IS. The SU(2) and SU(1,1) IS have been 
shown recently to be useful for improving the accuracy of 
interferometric measurements \cite{HiMl}. The investigation of the 
AES yields the most full information on the IS for generators of 
the corresponding Lie group. It is also possible to 
consider the states generated by the action of unitary group 
transformations on the IS.
The most convenient way to examine different subsets of the AES 
and relations between them is via the analytic 
representation of the AES in the standard coherent-state basis. 
In the present work the algebra-eigenstate method is applied to 
the two-photon group $H_{6}$ whose unitary transformations 
squeeze and displace single-mode photon states.

\section{The Fock-Bargmann representation of the two-photon algebra 
eigenstates}

The theoretical analysis \cite{SS,SS1} and experimental realization
\cite{SSe1,SSe2,SSe3} of squeezed states continue to attract 
considerable attention \cite{SS2}. Much of the work so far was 
concerned with the single-mode case whose group-theoretic basis lies 
in the two-photon Lie group $H_{6}$ \cite{Gil:rev}. The 
corresponding Lie algebra is spanned by the six operators
$\{N, a^{2}, a^{\dagger 2}, a, a^{\dagger}, I \}$,
	\begin{equation}
\begin{array}{c}
{[a^{2},a^{\dagger 2}] = 4N + 2I} , \;\;\;\;\;\;\;\;
{[a,a^{\dagger}] = I} , \\
{[a^{\dagger 2},a] = -2 a^{\dagger}} , \;\;\;\;\;\;\;\;
{[a^{2},a^{\dagger}] = 2 a} , \\
{[N,a^{\dagger 2}] =  2 a^{\dagger 2}} , \;\;\;\;\;\;\;\;
{[N,a^{2}] = -2 a^{2}} , \\
{[N,a^{\dagger}] = a^{\dagger}} , \;\;\;\;\;\;\;\;
{[N,a] = - a} , 
\end{array}         \label{3.1}
	\end{equation}
where $N = a^{\dagger} a$ is the number operator. All the other
commutation relations are zero. The unified group-theoretic 
description of various types of states associated with the 
$H_{6}$ transformations can be obtained by means of the 
algebra-eigenstate method. This provides the analytic 
representation of single-mode photon states generated by squeezing 
and displacement group operators.

The representation Hilbert space of $H_{6}$ is the Fock space 
of the quantum harmonic oscillator. The orthonormal basis in this 
space is the Fock basis of the number eigenstates $|n\rangle$ 
$(n=0,1,\ldots,\infty)$. For any Fock state $|n\rangle$, the 
isotropy subgroup is U(1)$\otimes$U(1) with the algebra  spanned
by $\{ N,I \}$. The isotropy subgroup consists of all group elements 
$h$ of the form $h = \exp(i\delta N + i\varphi I)$. Thus $h|n\rangle 
= \exp(i\delta n + i\varphi ) |n\rangle$. The oscillator group
$H_{4}$ is a subgroup of $H_{6}$. The corresponding 
solvable Lie algebra is spanned by the four operators 
$\{N,a,a^{\dagger},I \}$. The quotient space 
$H_{4}/$U(1)$\otimes$U(1) can be parametrized by 
an arbitrary complex number $\alpha$. Then an element $\Omega\in 
H_{4}/$U(1)$\otimes$U(1) can be written as the displacement 
operator, $\Omega \equiv D(\alpha) = \exp(\alpha a^{\dagger} - 
\alpha^{\ast} a)$. Note that the same quotient space and hence the 
same set of the CS is obtained also for the Heisenberg-Weyl group 
$H_{3}$. This is a subgroup of $H_{4}$ 
($H_{3} \subset H_{4} \subset H_{6}$), and the 
nilpotent Lie algebra corresponding to  $H_{3}$ is spanned by 
the three operators $\{ a, a^{\dagger}, I \}$. 
The quotient space $H_{3}/$U(1) is the same as the space 
$H_{4}/$U(1)$\otimes$U(1).

The standard Glauber set of the CS is obtained when the vacuum 
state $|0\rangle$ is chosen as the reference state,
	\begin{equation}
|\alpha\rangle = D(\alpha) |0\rangle  = e^{-|\alpha|^{2}/2}  
\sum_{n=0}^{\infty} \frac{\alpha^{n}}{\sqrt{n!}} |n\rangle .  
\label{3.2}
	\end{equation}
For any state $|\Psi\rangle = \sum_{n=0}^{\infty} c_{n} |n\rangle$ 
in the Hilbert space, one can construct the entire analytic 
function \cite{FB}
	\begin{equation}
f(\alpha) = e^{|\alpha|^{2}/2} \langle \alpha^{\ast} | \Psi \rangle 
= \sum_{n=0}^{\infty} c_{n} \frac{\alpha^{n}}{\sqrt{n!}} . 
\label{3.3}
	\end{equation}
Then the identity resolution,
$(1/\pi) \int d^{2}\! \alpha \, 
|\alpha\rangle \langle\alpha| = I$,
can be used to expand the state $|\Psi\rangle$ in the 
coherent-state basis:
	\begin{equation}
|\Psi\rangle = \frac{1}{\pi} \int d^{2}\! \alpha \, 
e^{-|\alpha|^{2}/2} f(\alpha^{\ast}) |\alpha\rangle .  \label{3.5}
	\end{equation}
It is customary in quantum mechanics to restrict the Hilbert space 
to consist of normalizable states that satisfy the condition
	\begin{equation}
\langle \Psi | \Psi \rangle = \frac{1}{\pi} \int d^{2}\! \alpha \, 
e^{-|\alpha|^{2}} |f(\alpha^{\ast})|^{2} < \infty .   \label{3.6}
	\end{equation}
The analytic representation (\ref{3.3}) is known as the 
Fock-Bargmann representation \cite{FB}. The Glauber coherent state 
$|\upsilon\rangle$ is represented by 
the function
	\begin{equation}
{\cal F}(\upsilon;\alpha) = e^{|\alpha|^{2}/2} 
\langle \alpha^{\ast} | \upsilon \rangle
 = e^{-|\upsilon|^{2}/2} e^{\upsilon\alpha}  .  \label{3.7}
	\end{equation}
The generators of $H_{6}$ act in the Hilbert space of entire
analytic functions $f(\alpha)$ as linear differential operators:
	\begin{equation}
\begin{array}{c}
\vspace{0.2cm}
a = \displaystyle{ \frac{d}{d\alpha} } , \;\;\;\;\;\;\;\;
a^{\dagger} = \alpha , \;\;\;\;\;\;\;\; I = 1 , \\
N = \alpha \displaystyle{ \frac{d}{d\alpha} } , \;\;\;\;\;\;\;\;
a^{2} = \displaystyle{ \frac{d^{2}}{d\alpha^{2}} } , 
\;\;\;\;\;\;\;\;  a^{\dagger 2} = \alpha^{2} .
\end{array}   \label{3.8}
	\end{equation}

The two-photon AES are determined by the eigenvalue equation
	\begin{equation}
( \beta_{1} N + \beta_{2} a^{2} + \beta_{3} a^{\dagger 2} + 
\beta_{4} a + \beta_{5} a^{\dagger} ) |\lambda,\bbox{\beta}\rangle 
= \lambda |\lambda,\bbox{\beta}\rangle .   \label{3.9}
	\end{equation}
The AES $|\lambda,\bbox{\beta}\rangle$ are represented by the 
function
	\begin{equation}
\Lambda(\lambda,\bbox{\beta};\alpha) = e^{|\alpha|^{2}/2} 
\langle \alpha^{\ast} | \lambda,\bbox{\beta} \rangle , \label{3.10}
	 \end{equation}
and in the Fock-Bargmann representation the eigenvalue equation 
(\ref{3.9}) becomes  the second-order linear homogeneous 
differential equation
	\begin{equation}
\beta_{2} \frac{d^{2} \Lambda}{d \alpha^{2}} + ( \beta_{1} \alpha 
+ \beta_{4} ) \frac{d\Lambda}{d\alpha} + ( \beta_{3} \alpha^{2} + 
\beta_{5} \alpha - \lambda ) \Lambda = 0 .   \label{3.11}
	\end{equation}
By using the transformation 
	\begin{equation}
\Lambda(\lambda,\bbox{\beta};\alpha) = \exp \left( \frac{ 
\Delta -\beta_{1} }{ 4\beta_{2} } \alpha^{2} \right)
T(\lambda,\bbox{\beta};\alpha) ,   \label{3.12}
	\end{equation}
we get the equation with coefficients that are linear 
in $\alpha$,
	\begin{equation}
\beta_{2} \frac{d^{2} T}{d \alpha^{2}} + ( \Delta \alpha + 
\beta_{4} ) \frac{d T}{d\alpha} + \left[ \sigma \alpha + 
\mbox{\small{$\frac{1}{2}$}} (\Delta-\beta_{1}) - \lambda \right] T 
= 0 ,   \label{3.13}
	\end{equation}
where 
	\begin{eqnarray}
& & \Delta^{2} \equiv \beta_{1}^{2}-4\beta_{2}\beta_{3} ,
\label{3.14} \\
& & \sigma \equiv \beta_{4}\frac{\Delta-\beta_{1}}{2\beta_{2}} + 
\beta_{5} . \label{3.15}
	\end{eqnarray}
Note the double-valuedness of $\Delta$. Equation (\ref{3.13}) can 
be transformed into the Kummer equation for the confluent 
hypergeometric function or into the Bessel equation, depending on 
values of the parameters \cite{Erd}.

In the most general case, $\beta_{2} \neq 0$, $\Delta \neq 0$, two
independent solutions of Eq.\ (\ref{3.13}) are given by \cite{Erd}
\begin{mathletters}
\label{3.16}
	\begin{eqnarray}
& & T_{1}(\lambda,\bbox{\beta};\alpha) = \exp\left( 
-\frac{\sigma}{\Delta} \alpha \right) {}_{1}\! F_{1} \left( d 
\left| \frac{1}{2} \left| -\frac{\Delta}{2\beta_{2}} 
(\alpha - \mu_{\Delta})^{2} \right. \right. \right) ,  
\label{3.16a}  \\
& & T_{2}(\lambda,\bbox{\beta};\alpha) = \sqrt{
-\frac{\Delta}{2\beta_{2}}}\, (\alpha - \mu_{\Delta}) 
\exp\left( -\frac{\sigma}{\Delta} \alpha \right) 
{}_{1}\! F_{1} \left( d + \frac{1}{2} \left| \frac{3}{2}  
\left| -\frac{\Delta}{2\beta_{2}} (\alpha - \mu_{\Delta})^{2} 
\right. \right. \right) ,  \label{3.16b}
	\end{eqnarray}
\end{mathletters}
where 
	\begin{equation}
\mu_{\Delta} \equiv (2\beta_{2}\beta_{5} - \beta_{1}\beta_{4})/ 
\Delta^{2},  \label{3.17}
	\end{equation}
	\begin{equation}
d \equiv \frac{1}{2\Delta} \left( \beta_{2} \frac{\sigma^{2}
}{\Delta^{2}} - \beta_{4} \frac{\sigma}{\Delta} + 
\frac{\Delta - \beta_{1}}{2} - \lambda \right) , \label{3.18}
	\end{equation}
and ${}_{1}\! F_{1} (d|c|x)$ is the confluent hypergeometric 
function (the Kummer function). Note that the function 
${}_{1}\! F_{1} (d|c|x)$ with $c=1/2$ or $c=3/2$ can be expressed
in terms of the parabolic cylinder functions $D_{\nu}(\pm x)$ by 
using the relation \cite{AS}
	\begin{equation}
D_{\nu}(\pm x) = \sqrt{\pi}\, 2^{\nu/2} e^{-x^{2}/4} \left[
\frac{1}{ \Gamma\left(\frac{1-\nu}{2}\right) }\, {}_{1}\! F_{1} 
\left(-\frac{\nu}{2} \left| \frac{1}{2} \left|
\frac{x^{2}}{2} \right. \right. \right)  
\mp \frac{\sqrt{2}\,x}{\Gamma\left(-\frac{\nu}{2}
\right)}\, {}_{1}\! F_{1} \left(\frac{1-\nu}{2} \left| \frac{3}{2} 
\left|\frac{x^{2}}{2}\right. \right. \right) \right] . \label{pcf}
	\end{equation}
The function $\Lambda(\lambda,\bbox{\beta};\alpha)$ is manifestly 
analytic, and the normalization condition (\ref{3.6})
requires 
	\begin{equation}
\left| \frac{\Delta \pm \beta_{1}}{2\beta_{2}} \right| < 1 . 
\label{3.19}
	\end{equation}
The sign `$-$' in Eq.\ (\ref{3.19}) must be taken when $d$ 
($d+\frac{1}{2}$) is a nonpositive integer and the sign `$+$' 
otherwise. 

The physical meaning of the two solutions can be understood by 
considering the particular case $\beta_{4} = \beta_{5} =0$ when
one-photon processes are excluded. Then $\sigma = 0$, and
$\mu_{\Delta} = 0$, so $T_{1}(\lambda,\bbox{\beta};\alpha)$ 
contains only even powers of $\alpha$ and 
$T_{2}(\lambda,\bbox{\beta};\alpha)$ contains only odd powers of 
$\alpha$. If we recall that the operators $\{ N, a^{2}, 
a^{\dagger 2} \}$ form a realization of the SU(1,1) Lie algebra, 
it will be clear that the two solutions represent the AES in the 
two irreducible sectors of SU(1,1). One-photon processes 
represented by $a$ and $a^{\dagger}$ mix these irreducible sectors, 
and then the total solution is given by a superposition of $T_{1}$
and $T_{2}$.

In the degenerate case $\Delta = 0$. Provided that 
$\beta_{2} \neq 0$, $\sigma \neq 0$, a solution
of Eq.\ (\ref{3.13}) is given by \cite{Erd}
	\begin{equation}
T(\lambda,\bbox{\beta};\alpha) = \exp\left( -\frac{\beta_{4}
}{2\beta_{2}} \alpha \right) \sqrt{\alpha-\mu_{0}} \,\,
J_{1/3} \left( \frac{2}{3} \sqrt{\frac{\sigma}{
\beta_{2}}} (\alpha-\mu_{0})^{3/2} \right) ,  \label{3.20}
	\end{equation}
where
	\begin{equation}
\mu_{0} \equiv \frac{ 4\beta_{2}\lambda + 2\beta_{1}\beta_{2}
+ \beta_{4}^{2} }{ 4\beta_{2}\sigma } ,  \label{3.21}
	\end{equation}
and $J_{\nu}(x)$ is the Bessel function of the first kind. Another
independent solution includes $J_{-\nu}(x)$ instead of $J_{\nu}(x)$
(for a noninteger $\nu$). The solution can be also expressed in 
terms of the Airy functions \cite{AS}:
\begin{mathletters} \label{airy}
	\begin{eqnarray}
& & {\rm Ai}(-x) = \frac{\sqrt{x}}{3} \left[J_{1/3}
\left( \frac{2}{3} x^{3/2}\right) + J_{-1/3}
\left(\frac{2}{3} x^{3/2} \right) \right] , \label{airy_a} \\
& & {\rm Bi}(-x) = \sqrt{ \frac{x}{3} } \left[ J_{-1/3}
\left(\frac{2}{3} x^{3/2}\right) - J_{1/3}
\left(\frac{2}{3} x^{3/2}\right) \right] . \label{airy_b}
	\end{eqnarray}
\end{mathletters}
The function $\Lambda(\lambda,\bbox{\beta};\alpha)$ is manifestly 
analytic, and the normalization condition (\ref{3.6}) requires 
$|\beta_{1}/\beta_{2}| < 2$. 

When, apart from $\Delta = 0$, also $\sigma=0$, then Eq.\ 
(\ref{3.13}) becomes an equation with constant coefficients. 
The solution of this equation can be written in terms of elementary
functions:
	\begin{equation}
T(\alpha) = C_{+} \exp(\omega_{+}\alpha) + 
C_{-} \exp(\omega_{-}\alpha) ,   \label{lin1} 
	\end{equation}
where $C_{\pm}$ are the integration constants, and
	\begin{equation}
\omega_{\pm} = \frac{1}{2\beta_{2}} \left( -\beta_{4} \pm 
\sqrt{ \beta_{4}^{2} + 4\beta_{2}\lambda +2\beta_{2}\beta_{2} }
\right) .  \label{lin2}
	\end{equation}

In the case $\beta_{2} = 0$, $\beta_{1} \neq 0$, the eigenvalue 
equation (\ref{3.11}) is a first-order differential equation whose 
solution is easily found to be
	\begin{equation}
\Lambda(\lambda,\bbox{\beta};\alpha) = \Lambda_{0} \left( 
\alpha + \frac{\beta_{4}}{\beta_{1}} \right)^{p}
\exp\left( -\frac{\beta_{3}}{2\beta_{1}} \alpha^{2} + 
\frac{ \beta_{3}\beta_{4} - \beta_{1}\beta_{5} }{ \beta_{1}^{2} } 
\alpha \right)  ,  \label{3.23}
	\end{equation}
where 
	\begin{equation}
p = [\beta_{1}^{2}\lambda - \beta_{4}(\beta_{3}\beta_{4} 
- \beta_{1}\beta_{5})]/\beta_{1}^{3}
	\end{equation}
must be a non-negative integer in order to satisfy the analyticity
condition. $\Lambda_{0}$ is a normalization factor, and 
the normalization condition (\ref{3.6}) requires 
$|\beta_{3}/\beta_{1}| < 1$. 

When $\beta_{2} = 0$ and $\beta_{3} = 0$, the resulting AES are 
associated with the oscillator group $H_{4}$. 
The corresponding analytic function is
	\begin{equation}
\Lambda(\lambda,\bbox{\beta};\alpha) = \Lambda_{0} \left( \alpha + 
\frac{\beta_{4}}{\beta_{1}} \right)^{p} \exp\left( 
-\frac{\beta_{5}}{\beta_{1}} \alpha \right)  ,  \label{3.25}
	\end{equation}
where
	\begin{equation}
p = (\beta_{1}\lambda + \beta_{4}\beta_{5})/\beta_{1}^{2} 
\label{3.26}
	\end{equation}
is once again a non-negative integer. The function
$\Lambda(\lambda,\bbox{\beta};\alpha)$ of Eq.\ (\ref{3.25}) 
represents displaced Fock states \cite{DFS}. In order to derive the
corresponding eigenvalue equation, we start from the equation
$N|n\rangle = n|n\rangle$. By applying the unitary displacement 
operator $D(\upsilon) = \exp(\upsilon a^{\dagger} - 
\upsilon^{\ast} a)$ to this equation, we obtain
	\begin{equation}
(N - \upsilon^{\ast} a - \upsilon a^{\dagger}) |n,\upsilon\rangle
= (n - |\upsilon|^{2}) |n,\upsilon\rangle ,  \label{3.27}
	\end{equation}
where $|n,\upsilon\rangle = D(\upsilon) |n\rangle$ is the displaced
Fock state that reduces to the standard Glauber state for $n=0$.
The corresponding analytic function is given by Eq.\ (\ref{3.25}).
By substituting 
	\begin{equation}
\beta_{1} = 1, \;\;\;\;\;\;\; 
\beta_{5} = \beta_{4}^{\ast} = -\upsilon, \;\;\;\;\;\;\;
\lambda = n - |\upsilon|^{2} ,   \label{3.28}
	\end{equation}
we find $p=n$, and
	\begin{equation}
\Lambda(n,\upsilon;\alpha) = \Lambda_{0} ( \alpha - 
\upsilon^{\ast} )^{n} e^{\upsilon\alpha} .  \label{3.29}
	\end{equation}
For $n=0$, this function reduces to the function 
${\cal F}(\upsilon;\alpha)$ representing the Glauber CS 
$|\upsilon\rangle$ [cf. Eq.\ (\ref{3.7})]. The normalization 
factor in this case is $\Lambda_{0} = \exp(-|\upsilon|^{2}/2)$.
A consequence of Eq.\ (\ref{3.27}) is the following equation
satisfied by the Glauber CS
	\begin{equation}
(N - \upsilon^{\ast} a - \upsilon a^{\dagger} + |\upsilon|^{2}) 
|\upsilon\rangle = 0 .   \label{3.30}
	\end{equation}
By using the Glauber definition $a |\upsilon\rangle = \upsilon
|\upsilon\rangle$, we see that Eq.\ (\ref{3.30}) is an identity.

We also consider another example of displaced states. In the case 
$\beta_{1} = \beta_{2} = \beta_{3} = 0$, $\beta_{4} \neq 0$, the
resulting AES are associated with the $H_{3}$ group. Then the 
solution of the eigenvalue equation (\ref{3.11}) is
	\begin{equation}
\Lambda(\lambda,\bbox{\beta};\alpha) = \Lambda_{0} 
\exp\left( - \frac{\beta_{5}}{2\beta_{4}} \alpha^{2}
+  \frac{\lambda}{\beta_{4}} \alpha \right) .  \label{3.31}
	\end{equation}
We see that the analyticity condition is fulfilled. Besides, the 
normalization condition (\ref{3.6}) requires $|\beta_{5}/\beta_{4}|
< 1$. By comparing the function 
$\Lambda(\lambda,\bbox{\beta};\alpha)$ of Eq.\ (\ref{3.31}) with 
the function ${\cal F}(\upsilon;\alpha)$ of Eq.\ (\ref{3.7}), we 
find that the algebra eigenstate $|\lambda,\bbox{\beta}\rangle$ 
coincides with the Glauber coherent state $|\upsilon\rangle$ for 
$\beta_{i} = 0$, $i=1,2,3,5$. Then $\upsilon = \lambda/\beta_{4}$,
and Eq.\ (\ref{3.9}) reduces to the famous Glauber equation 
$a |\upsilon\rangle = \upsilon|\upsilon\rangle$. We see that the
eigenvalue equation for a state (e.g., for the standard coherent 
state) can be written in a number of ways, i.e., there is a number 
of equivalent definitions of the state. Note also that in the case 
$\beta_{i} = 0$, $i=1,2,3,4$, $\beta_{5} \neq 0$, Eq.\ (\ref{3.9}) 
has not any nontrivial solution. The reason is that the creation 
operator $a^{\dagger}$ has not any eigenstate.

The Gaussian form of the function $\Lambda(\lambda,\bbox{\beta};
\alpha)$ of Eq.\ (\ref{3.31}) means that this function represents 
displaced (canonical) squeezed states of Stoler and Yuen \cite{SS}. 
These states are generated by the action of the squeezing and 
displacement operators on the vacuum \cite{SS},
	\begin{equation}
|\xi,\upsilon\rangle = D(\upsilon) S(\xi) |0\rangle ,  
\label{3.32}
	\end{equation}
where the squeezing operator is
	\begin{equation}
S(\xi) = \exp(\mbox{\small{$\frac{1}{2}$}}\xi a^{\dagger 2} 
- \mbox{\small{$\frac{1}{2}$}}\xi^{\ast} a^{2}) . 
\label{3.33}
	\end{equation}
By applying the squeezing operator $S(\xi = s\, e^{i\theta})$ to the 
equation $a|0\rangle = 0$, one derives the equation satisfied by the 
squeezed vacuum $|\xi\rangle = S(\xi) |0\rangle$,
	 \begin{equation}
[ (\cosh s)a - (\sinh s \, e^{i\theta}) a^{\dagger}] |\xi\rangle 
= 0 .  \label{3.34}
	\end{equation}
By applying the displacement operator $D(\upsilon)$ to this 
equation, one finds the eigenvalue equation satisfied by the 
displaced squeezed state $|\xi,\upsilon\rangle$,
	\begin{equation}
( a - \zeta a^{\dagger}) |\xi,\upsilon\rangle = (\upsilon - 
\zeta \upsilon^{\ast}) |\xi,\upsilon\rangle ,  \label{3.35}
	\end{equation}
where
	\begin{equation}
\zeta \equiv \frac{\xi}{|\xi|} \tanh |\xi| 
= \tanh s \, e^{i\theta} .  \label{3.zeta}
	\end{equation}
By substituting
	\begin{equation}
\beta_{4} = 1 , \;\;\;\;\;\;\;\; 
\beta_{5} = - \zeta , \;\;\;\;\;\;\;\;
\lambda = \upsilon - \zeta \upsilon^{\ast}    \label{3.36}
	\end{equation}
into Eq.\ (\ref{3.31}), one obtains the analytic function 
representing the displaced squeezed states,
	\begin{equation}
\Lambda(\xi,\upsilon;\alpha) = \Lambda_{0} \exp\left[ 
\mbox{\small{$\frac{1}{2}$}} \zeta \alpha^{2} +
(\upsilon - \zeta \upsilon^{\ast}) \alpha \right] .  \label{3.37}
	\end{equation}
The normalization factor in this case is \cite{SS}
	\begin{equation}
\Lambda_{0} = \frac{ \exp( -\frac{1}{2}|u|^{2} 
- \zeta^{\ast} u^{2} ) }{\sqrt{\cosh s}}, 
	\end{equation}
where 
	\begin{equation}
u \equiv (\cosh s)\, \upsilon - \left(\sinh s\, e^{i\theta} 
\right) \upsilon^{\ast} .  \label{3.u}
	\end{equation}
It is interesting to note that the displaced squeezed states 
$|\xi,\upsilon\rangle$ are the standard CS of the group 
$H_{6}$ but simultaneously they are nonstandard CS of its 
subgroup $H_{3}$. 
The reference state of this nonstandard set is the squeezed vacuum 
$|\xi\rangle$. The displaced squeezed states $|\xi,\upsilon\rangle$
are also the generalized IS for the quadratures $X_{1}$ and $X_{2}$
that are the Hermitian generators of $H_{3}$. By putting
$a = X_{1} +iX_{2}$ and $a^{\dagger} = X_{1} -iX_{2}$ in the 
eigenvalue equation (\ref{3.35}), one obtains the equation of type
(\ref{2.15}):
	\begin{equation}
\left[\left(\frac{1-\zeta}{1+\zeta} \right) X_{1} +iX_{2} \right]
|\xi,\upsilon\rangle = \left(\frac{ \upsilon -\zeta\upsilon^{\ast}
}{1+\zeta} \right) |\xi,\upsilon\rangle .  \label{3.38}
	\end{equation}
(The $X_{1}$-$X_{2}$ generalized IS also are known as ``correlated
coherent states'' \cite{corst}).
For $\theta = 0$ and $\theta = \pi$, $\zeta$ is real and the
$|\xi,\upsilon\rangle$ states are the ordinary IS, i.e., they 
provide an equality in the uncertainty relation
$\Delta X_{1} \Delta X_{2}\geq 1/4$. 
The Glauber CS $|\upsilon\rangle$ form the zero-squeezing 
subset of the $X_{1}$-$X_{2}$ IS.

\section{Displaced and squeezed Fock states}

The differential equation  (\ref{3.11}) determines analytic 
functions representing various photon states that can be produced
by squeezing and displacement of an initial state. The first
candidate to be the initial state is the vacuum. Recently, have
been considerable interest in attempts to produce Fock states 
(photon number eigenstates) $|n\rangle$ with nonzero occupation 
number \cite{FSP:mm,FSP_SCP,FSP:pa,FSP:qj,FSP:rl,FSP:sai}. Given 
that Fock states can be generated, it is natural to consider their
displacement (by driving the light field by a classical current)
and squeezing (by degenerate parametric amplification). Properties 
of displaced Fock states \cite{DFS,KOK:jmo,DFS:jcm,DFS:gen,DFS:pp}, 
squeezed Fock states \cite{SFS,KOK:jmo,SFS:pp}, 
and displaced and squeezed Fock states (DSFS) 
\cite{Kral,Lo} have been widely discussed.
In this section we consider the DSFS as a characteristic example of 
the two-photon AES. The general results of the preceding section
are used to obtain the Fock-Bargmann analytic representation of the 
DSFS.  

We start from the equation $N|n\rangle = n|n\rangle$. By acting 
on both sides of this equation with the squeezing operator
$S(\xi=se^{i\theta})$, we derive the eigenvalue equation 
satisfied by the squeezed Fock states $|n,\xi\rangle = 
S(\xi)|n\rangle$,
	\begin{equation}
( \beta_{1} N + \beta_{2} a^{2} + \beta_{3} a^{\dagger 2} ) 
|n,\xi\rangle = (n-\sinh^{2}\! s) |n,\xi\rangle ,   \label{4.1}
	\end{equation}
where 
	\begin{equation}
\beta_{1} = \cosh 2s , \;\;\;\;\;\;\;
\beta_{2} = \beta_{3}^{\ast} = 
-\frac{1}{2} \sinh 2s\, e^{-i\theta} .  \label{4.2}
	\end{equation}
Then we apply the displacement operator $D(\upsilon=re^{i\phi})$.
The resulting eigenvalue equation reads
	\begin{equation}
( \beta_{1} N + \beta_{2} a^{2} + \beta_{3} a^{\dagger 2} 
+ \beta_{4} a + \beta_{5} a^{\dagger} ) |n,\xi,\upsilon\rangle = 
\lambda |n,\xi,\upsilon\rangle ,   \label{4.3}
	\end{equation}
where 
	\begin{equation}
|n,\xi,\upsilon\rangle = D(\upsilon) S(\xi) |n\rangle  \label{4.4}
	\end{equation}
are the DSFS. The parameters $\beta_{1}$, $\beta_{2}$ and 
$\beta_{3}$ remain as given above, and
	\begin{eqnarray}
& & \beta_{4} = \beta_{5}^{\ast} = \upsilon^{\ast}\left(
\sinh 2s\, e^{-i(\theta-2\phi)} - \cosh 2s \right) , \label{4.5} \\
& & \lambda = n-\sinh^{2}\! s + r^{2}[ \sinh 2s\, 
\cos (\theta-2\phi) - \cosh 2s] .     \label{4.6}
	\end{eqnarray}
These results can be easily derived by using the general recipe
	\begin{equation}
D(\upsilon) F(a,a^{\dagger}) D^{-1}(\upsilon) =
F(a-\upsilon,a^{\dagger}-\upsilon^{\ast}) ,   \label{4.7} 
	\end{equation}
where $F(a,a^{\dagger})$ is a power series. 

Equation (\ref{4.3}) is of the general form (\ref{3.9}) and the
corresponding differential equation is of the form (\ref{3.11})
with solutions given by Eqs.\ (\ref{3.12}) and (\ref{3.16}). 
A simple calculation yields
	\begin{equation}
\Delta^{2} = \beta_{1}^{2} -4\beta_{2}\beta_{3} = 1 ,  \label{4.8}
	\end{equation}
which is a direct consequence of the unitarity of the squeezing
operator $S(\xi)$. Then $\Delta = \pm 1$, and we find, respectively,
	\begin{equation}
\sigma = \pm \upsilon \left[ \left( \frac{\sinh s}{\cosh s} 
\right)^{\pm 1} e^{i(\theta-2\phi)} - 1 \right] ,  \;\;\;\;\;\;
\mu_{\Delta} = \upsilon^{\ast} ,    \;\;\;\;\;\;
d = \mp \frac{1}{2} \left( n +\frac{1}{2} \mp\frac{1}{2} \right) . 
\label{4.11}
	\end{equation}

Let us start from $\Delta =+1$. Then $d=-n/2$, and the normalization
condition (\ref{3.19}) is satisfied by taking $T_{1}(\lambda,
\bbox{\beta};\alpha)$ for even values of $n$ and $T_{2}(\lambda,
\bbox{\beta};\alpha)$ for odd values of $n$. This result is dictated
by the fact that the analytic function representing the squeezed
Fock states $|n,\xi\rangle$ contains only even powers of $\alpha$
for even $n$ and only odd powers of $\alpha$ for odd $n$. By using
the relations between the confluent hypergeometric functions and the 
Hermite polynomials,
\begin{mathletters} \label{4.12}
	\begin{eqnarray}
& &  {}_{1}\! F_{1} \left(-m \left| \frac{1}{2} \right| x^{2} 
\right) = \frac{(-1)^{m} m! H_{2m}(x)}{(2m)!} ,  \label{4.12a} \\
& &  x\, {}_{1}\! F_{1} \left(-m \left| \frac{1}{2} \right| x^{2} 
\right) = \frac{(-1)^{m} m! H_{2m+1}(x)}{2 (2m+1)!} ,   
\label{4.12b}
	\end{eqnarray}
\end{mathletters}
we find the solution:
	\begin{equation}
\Lambda(n,\xi,\upsilon;\alpha) = e^{|\alpha|^{2}/2}
\langle\alpha^{\ast}|n,\xi,\upsilon\rangle =
\Lambda_{0}(n,\xi,\upsilon) \exp\left[  \frac{\zeta}{2}
\alpha^{2} + (\upsilon-\zeta\upsilon^{\ast})\alpha \right] 
H_{n}\left(\frac{\alpha-\upsilon^{\ast}}{\sqrt{\sinh 2s\,
e^{-i\theta}}} \right) .   \label{4.13}
	\end{equation}
As usual, $\Lambda_{0}$ is a normalization factor, and $\zeta$ is
defined by Eq.\ (\ref{3.zeta}). This result is in accordance with
the expression for $\langle\alpha|S(\xi)D(\upsilon)|n\rangle$
derived in a different way by Kr\'{a}l \cite{Kral}.
The normalization factor is identified to be
	\begin{equation}
\Lambda_{0}(n,\xi,\upsilon) = \frac{ (\zeta^{\ast}/2)^{n/2} 
}{ \sqrt{n!\cosh s} } \exp\left( -\frac{1}{2}|u|^{2} - \zeta^{\ast} 
u^{2} \right) ,      \label{4.14}
	\end{equation}
where $u$ is defined by Eq.\ (\ref{3.u}). 

It is well known \cite{Erd} that the confluent hypergeometric 
function can be written in two equivalent forms which are related
by Kummer's transformation
	\begin{equation}
{}_{1}\! F_{1} \left(d \left| c \left| x \right. \right. \right)
= e^{x} {}_{1}\! F_{1} \left(c-d \left| c \left| -x 
\right. \right. \right) .   \label{4.15}
	\end{equation}
It is not difficult to see that the choice $\Delta = -1$ leads to
the solution which is related to the function 
$\Lambda(n,\xi,\upsilon;\alpha)$ of Eq.\ (\ref{4.13}) by Kummer's
transformation (\ref{4.15}). Then the solution can be written in
the form
\begin{mathletters}  \label{4.16}
	\begin{equation}
\Lambda_{1}(n,\xi,\upsilon;\alpha) = \Lambda_{0}^{(1)}
\exp\left[ \frac{\alpha^{2}}{2\zeta^{\ast}} + (\upsilon-
\upsilon^{\ast}/\zeta^{\ast})\alpha \right] 
{}_{1}\! F_{1} \left(\frac{n+1}{2} \left| \frac{1}{2} 
\left| \frac{ -(\alpha-\upsilon^{\ast})^{2} }{ \sinh 2s\, 
e^{-i\theta} } \right. \right. \right)   \label{4.16a}  
	\end{equation}
for even values of $n$ and
	\begin{equation}
\Lambda_{2}(n,\xi,\upsilon;\alpha) = \Lambda_{0}^{(2)}
\exp\left[ \frac{\alpha^{2}}{2\zeta^{\ast}} + (\upsilon-
\upsilon^{\ast}/\zeta^{\ast})\alpha \right] 
(\alpha-\upsilon^{\ast}) \,
{}_{1}\! F_{1} \left(\frac{n+2}{2} \left| \frac{3}{2} 
\left| \frac{ -(\alpha-\upsilon^{\ast})^{2} }{ \sinh 2s\, 
e^{-i\theta} } \right. \right. \right)   \label{4.16b}  
	\end{equation}
\end{mathletters}
for odd values of $n$. As usual, $\Lambda_{0}$ are appropriate
normalization factors, and the normalization condition (\ref{3.19})
is obviously satisfied.

In the particular case $n=0$, the function 
$\Lambda(n,\xi,\upsilon;\alpha)$ given by Eq.\ (\ref{4.13}) 
reduces to the function (\ref{3.37}) representing the displaced 
squeezed states.
The analytic function representing the squeezed Fock states
$|n,\xi\rangle$ is obtained by putting $\upsilon = 0$ in Eq.\
(\ref{4.13}) or in Eqs.\ (\ref{4.16}). The displaced Fock states
$|n,\upsilon\rangle$ were discussed in the preceding section and 
the corresponding analytic function is given by Eq.\ (\ref{3.29}).

We finish this section by a short review of basic methods for
producing the DSFS. Displacement can be implemented by linear
amplification of the light field. A usual method for doing that
is by driving the field by a classical current. The use of a
linear directional coupler as a displacing device was also
discussed \cite{LBK}. The most frequently used squeezing device
in the single-mode case is a degenerate parametric amplifier.
These methods of displacement and squeezing are well developed
and the main problem remaining is the production of a stable Fock 
state that will serve as the input state of displacing and 
squeezing devices. It was demonstrated that it is possible to 
generate a Fock state of the single-mode electromagnetic field 
in a micromaser operated under the appropriate conditions 
\cite{FSP:mm,FSP_SCP}. Another interesting method for producing 
Fock states is based on the process of parametric down-conversion 
in which one pump photon is destroyed and two correlated photons 
are simultaneously created, one in each of two distinct modes. The
state of one mode is then conditioned on the detection of photons
in the other mode \cite{FSP:pa}. It was also shown that a Fock
state can be generated by observation of quantum jumps in an ion 
trap \cite{FSP:qj}, by coupling a cavity to a single three-level
atom in a Raman lambda configuration \cite{FSP:rl}, and by using
the single-atom interference \cite{FSP:sai}.

\section{Squeezing and displacement of coherent superposition
states}

In this section we will consider squeezed and displaced 
superpositions of the Glauber CS $|\upsilon\rangle$ and 
$|-\upsilon\rangle$, which provide an interesting example of the 
two-photon AES. These states belong to the wide class of 
macroscopic quantum superpositions which are frequently referred 
to as the Schr\"{o}dinger-cat states \cite{Sch_cat}. 
Properties of different types of the Schr\"{o}dinger-cat states 
have been recently studied in a number of works 
\cite{EOCS,Hil87_89,YuSt86,XiGu,BVBK,KiBu,%
HaGe:qo,Bu:psa,HaGe:jmo,XWHM,DomJa}. The problem of the generation
of optical superposition states have drawn recently a lot of 
attention \cite{YuSt86,MiHo,MeTo,WoCar,GeHa,GaKn,LyGN,GeaB,SKK,%
Mey,FSP_SCP,DMBRH,BGK,qndm}. It was shown that the 
Schr\"{o}dinger-cat states can be produced in various nonlinear
processes \cite{YuSt86,MiHo,MeTo,WoCar,GeHa,GaKn,LyGN}, 
in field-atom interactions \cite{GeaB,SKK,Mey,FSP_SCP,DMBRH,BGK},
and in quantum nondemolition measurements \cite{qndm}.

We start from the coherent superposition state of the form
	\begin{equation}
|\upsilon,\tau,\varphi\rangle = {\cal N} \left( |\upsilon\rangle
+\tau e^{i\varphi} |-\upsilon\rangle \right) ,   \label{5.1}
	\end{equation}
where $|\upsilon\rangle$ and $|-\upsilon\rangle$ are the standard
Glauber CS, $\tau$ and $\varphi$ are real parameters, and
	\begin{equation}
{\cal N} = \left( 1+\tau^{2}+2\tau e^{-2|\upsilon|^{2}} \cos\varphi
\right)^{-1/2}     \label{5.2}
	\end{equation}
is the normalization factor. The analytic function
	\begin{equation}
\Lambda(\upsilon,\tau,\varphi;\alpha) = e^{|\alpha|^{2}/2}
\langle\alpha^{\ast}|\upsilon,\tau,\varphi\rangle  \label{5.3}
	\end{equation}
can be straightforwardly calculated:
	\begin{equation}
\Lambda(\upsilon,\tau,\varphi;\alpha) = {\cal N} 
e^{-|\upsilon|^{2}/2} \left( e^{\upsilon\alpha} + \tau e^{i\varphi}
e^{-\upsilon\alpha} \right) .   \label{5.4}
	\end{equation}
The superposition $|\upsilon,\tau,\varphi\rangle$ is a special
kind of the two-photon AES since it is the eigenstate of the 
operator $a^{2}$:
	\begin{equation}
a^{2} |\upsilon,\tau,\varphi\rangle =
\upsilon^{2} |\upsilon,\tau,\varphi\rangle .  \label{5.5}
	\end{equation}
In the case $\tau=0$, this state reduces to the Glauber coherent 
state $|\upsilon\rangle$.

Interesting superpositions are even and odd CS 
$|\upsilon\rangle_{e}$ and $|\upsilon\rangle_{o}$ \cite{EOCS}:
\begin{mathletters}  \label{5.6}
	\begin{eqnarray}
|\upsilon\rangle_{e} & = & |\upsilon,\tau=1,\varphi=0\rangle =
\frac{ |\upsilon\rangle + |-\upsilon\rangle }{ \sqrt{ 2\left(
1 + e^{-2|\upsilon|^{2}} \right) } } ,  \label{5.6a} \\
|\upsilon\rangle_{o} & = & |\upsilon,\tau=1,\varphi=\pi\rangle =
\frac{ |\upsilon\rangle - |-\upsilon\rangle }{ \sqrt{ 2\left(
1 - e^{-2|\upsilon|^{2}} \right) } } .  \label{5.6b}
	\end{eqnarray}
\end{mathletters}
The even and odd CS have a number of interesting nonclassical 
properties. The even CS are highly squeezed in the field quadrature
$X_{2}$, while the odd CS have sub-Poissonian photon statistics
\cite{XiGu,BVBK}. Multimode versions of the even and odd CS
have been recently studied \cite{AM_DMN}.

In the case $\tau=1$, $\varphi=\pi/2$, one
obtains the so-called Yurke-Stoler state 
	\begin{equation}
|\upsilon\rangle_{\text{YS}} = \frac{1}{\sqrt{2}} \left(
|\upsilon\rangle + i|-\upsilon\rangle \right) ,  \label{5.7}
	\end{equation}
that can be generated when the Glauber state $|\upsilon\rangle$
propagates through a nonlinear Kerr medium \cite{YuSt86}. The
$|\upsilon\rangle_{\text{YS}}$ states are squeezed in the $X_{2}$
field quadrature \cite{BVBK}.

It follows from the eigenvalue equation (\ref{5.5}) that the 
superpositions $|\upsilon,\tau,\varphi\rangle$ are a special
case of the two-photon IS. More precisely, let us consider the
two-photon realization of the SU(1,1) Lie algebra:
	\begin{equation}
K_{+} = \frac{1}{2}a^{\dagger 2}, \;\;\;\;\;\;
K_{-} = \frac{1}{2}a^{2}, \;\;\;\;\;\;
K_{0} = \frac{1}{2}N + \frac{1}{4} ,  \label{5.8}
	\end{equation}
	\begin{equation}
[K_{-},K_{+}] = 2K_{0}, \;\;\;\;\;\; [K_{0},K_{\pm}] = \pm K_{\pm} .
\label{5.9}
	\end{equation}
It is clear that SU(1,1)$\,\subset H_{6}$. One can use the
Hermitian combinations
	\begin{equation}
\begin{array}{l} \vspace{0.2cm}
K_{1} = \displaystyle{ \frac{1}{2} } (K_{+} + K_{-}) = 
\displaystyle{ \frac{1}{4} } (a^{\dagger 2} + a^{2}) , \\
K_{2} = \displaystyle{ \frac{1}{2i} } (K_{+} - K_{-}) = 
\displaystyle{ \frac{1}{4i} } (a^{\dagger 2} - a^{2}) , \label{5.10}
\end{array}
	\end{equation}
which satisfy the commutation relation
$[K_{1},K_{2}] = -iK_{0}$.
According to the general formalism of section II D, the 
$|\upsilon,\tau,\varphi\rangle$ states are the $K_{1}$-$K_{2}$ IS,
i.e., they provide an equality in the uncertainty relation 
	\begin{equation}
(\Delta K_{1})^{2} (\Delta K_{2})^{2} \geq \frac{1}{4} 
\langle K_{0} \rangle^{2} .   \label{5.12}
	\end{equation}
Indeed, a simple calculation yields 
	\begin{equation}
(\Delta K_{1})^{2} = (\Delta K_{2})^{2} = \frac{1}{2}
\langle K_{0} \rangle =
\frac{|\upsilon|^{2}}{4} \frac{ 1 + \tau^{2} - 2\tau 
e^{-2|\upsilon|^{2}} \cos\varphi }{ 1 + \tau^{2} + 2\tau 
e^{-2|\upsilon|^{2}} \cos\varphi } + \frac{1}{8} ,  \label{5.13}
	\end{equation}
when the expectation values are calculated for the superpositions
$|\upsilon,\tau,\varphi\rangle$. 

Now, let us recall that the Barut-Girardello states are defined as
the eigenstates of the SU(1,1) lowering generator $K_{-}$ \cite{BG}.
For each unitary irreducible representation of SU(1,1), there is a
set of the Barut-Girardello states. In the case of the two-photon
realization (\ref{5.8}), there are two irreducible representations
and the two irreducible sectors are spanned by the Fock states
$|n\rangle$ with even and odd values of $n$, respectively. The
two sets of the Barut-Girardello states are the even and odd CS
$|\upsilon\rangle_{e}$ and $|\upsilon\rangle_{o}$. Their intelligent 
properties were first recognized by Hillery \cite{Hil87_89}.

Nonclassical properties of displaced even and odd CS were briefly
discussed by Xia and Guo \cite{XiGu}. Squeezed coherent 
superpositions were considered recently by Hach and Gerry 
\cite{HaGe:jmo} and by Xin {\em et al.} \cite{XWHM}. We will use
the algebra-eigenstate method developed above in order to obtain 
the Fock-Bargmann analytic representation of the squeezed and
displaced superpositions. By applying the squeezing operator 
$S(\xi=s e^{i\theta})$ to Eq.\ (\ref{5.5}), we find that the
squeezed superpositions
	\begin{equation}
|\upsilon,\tau,\varphi,\xi\rangle = S(\xi) 
|\upsilon,\tau,\varphi\rangle   \label{5.14}
	\end{equation}
satisfy the following eigenvalue equation
	\begin{equation}
a_{\xi}^{2} |\upsilon,\tau,\varphi,\xi\rangle
= \upsilon^{2} |\upsilon,\tau,\varphi,\xi\rangle ,  \label{5.15}
	\end{equation}
where 
	\begin{equation}
a_{\xi} = S(\xi) a S^{-1}(\xi) = (\cosh s) a - 
(\sinh s\, e^{i\theta}) a^{\dagger} .   \label{5.16}
	\end{equation}
Equation (\ref{5.15}) can be written in the standard form 
(\ref{3.9}):
	\begin{equation}
(-2\zeta N + a^{2} + \zeta^{2} a^{\dagger 2})
|\upsilon,\tau,\varphi,\xi\rangle 
= [\upsilon^{2} (1-|\zeta|^2)
+\zeta] |\upsilon,\tau,\varphi,\xi\rangle ,  \label{5.17}
	\end{equation}
where $\zeta = \tanh s\, e^{i\theta}$ is defined by Eq.\ 
(\ref{3.zeta}). Now, we apply the displacement operator $D(z)$.
The resulting eigenvalue equation is 
	\begin{equation}
a_{\xi,z}^{2} |\upsilon,\tau,\varphi,\xi,z\rangle
= \upsilon^{2} |\upsilon,\tau,\varphi,\xi,z\rangle ,  \label{5.18}
	\end{equation}
where 
	\begin{equation}
|\upsilon,\tau,\varphi,\xi,z\rangle = D(z) S(\xi) 
|\upsilon,\tau,\varphi\rangle   \label{5.19}
	\end{equation}
is the displaced and squeezed superposition, and
	\begin{equation}
a_{\xi,z} = D(z) S(\xi) a S^{-1}(\xi) D^{-1}(z)
= (\cosh s) (a-z) - (\sinh s\, e^{i\theta}) 
(a^{\dagger}-z^{\ast}) .     \label{5.20}
	\end{equation}
The standard form (\ref{3.9}) of the eigenvalue equation is
	\begin{equation}
(-2\zeta N + a^{2} + \zeta^{2} a^{\dagger 2} -2\rho a  
+2\xi\rho a^{\dagger}) |\upsilon,\tau,\varphi,\xi,z\rangle
= [\upsilon^{2} (1-|\zeta|^2) +\zeta -\rho^{2}] 
|\upsilon,\tau,\varphi,\xi,z\rangle ,  \label{5.21}
	\end{equation}
where we have defined
	\begin{equation}
\rho \equiv z - \zeta z^{\ast} .   \label{5.22}
	\end{equation}

A simple calculation yields 
	\begin{equation}
\Delta^{2} = \beta_{1}^{2} - 4\beta_{2}\beta_{3} = 0 , \;\;\;\;\;\;\;\;
\sigma = \beta_{4} \displaystyle{ \frac{ \Delta-\beta_{1}
}{ 2\beta_{2} } } + \beta_{5} = 0 .   \label{5.23}
	\end{equation}
The solution in this case is given by Eq.\ (\ref{lin1}). The
analytic function 
	\begin{equation}
\Lambda(\upsilon,\tau,\varphi,\xi,z;\alpha) = e^{|\alpha|^{2}/2}
\langle\alpha^{\ast}|\upsilon,\tau,\varphi,\xi,z\rangle  
\label{5.24}
	\end{equation}
is then given by 
	\begin{equation}
\Lambda(\upsilon,\tau,\varphi,\xi,z;\alpha) = \exp\left(
\frac{\zeta}{2} \alpha^{2} +\rho\alpha \right) \left[ 
C_{+} \exp\left( \frac{\upsilon\alpha}{\cosh s} \right)
+ C_{-} \exp\left(- \frac{\upsilon\alpha}{\cosh s}
\right) \right] .    \label{5.25}
	\end{equation}
This function is manifestly analytic and normalizable due to the
condition $|\zeta| < 1$. By putting $\rho=0$ in Eq.\ (\ref{5.25}),
we obtain the function that represents squeezed superpositions
$|\upsilon,\tau,\varphi,\xi\rangle$. The case of zero squeezing
is also included in Eq.\ (\ref{5.25}). By putting there $\zeta=0$,
we find the function that represents displaced superpositions
$|\upsilon,\tau,\varphi,z\rangle$.

By comparing the function 
$\Lambda(\upsilon,\tau,\varphi,\xi,z;\alpha)$ of Eq.\ (\ref{5.25})
with the function $\Lambda(\upsilon,\tau,\varphi;\alpha)$ of Eq.\
(\ref{5.4}), we deduce that
$C_{-} = \tau e^{i\varphi} C_{+}$, and 
$C_{+} = {\cal N} e^{-|\upsilon|^{2}/2}$ for $\zeta = z = 0$.
By using the generating function for the Hermite polynomials
\cite{Erd}
	\begin{equation}
e^{2tx-x^{2}} = \sum_{n=0}^{\infty} H_{n}(t) \frac{ x^{n} }{n!} ,
\label{5.28}
	\end{equation}
we expand the function $\Lambda(\upsilon,\tau,\varphi,\xi,z;\alpha)$
of Eq.\ (\ref{5.25}) into the power series in $\alpha$ and obtain
the Fock-state expansion of the displaced and squeezed 
superpositions:
	\begin{equation}
|\upsilon,\tau,\varphi,\xi,z\rangle = C_{+} \sum_{n=0}^{\infty}
\frac{ (-\zeta/2)^{n/2} }{ \sqrt{n!} } \left[ H_{n}\left(
\frac{u+\upsilon}{\kappa} \right) 
+ \tau e^{i\varphi} H_{n}\left(
\frac{u-\upsilon}{\kappa} \right) \right] |n\rangle ,  \label{5.29}
	\end{equation}
where we have defined 
	\begin{eqnarray}
& & u \equiv \rho \cosh s = 
\frac{z-\zeta z^{\ast}}{\sqrt{1-|\zeta|^{2}}} ,   \label{5.30} \\
& & \kappa \equiv i \sqrt{2\zeta} \, \cosh s
= i \sqrt{ \sinh 2s\, e^{i\theta} } .  \label{5.31}
	\end{eqnarray}
By using the summation theorem for Hermite polynomials \cite{Erd},
we readily find the normalization factor:
	\begin{equation}
C_{+}^{-2} = \frac{ \exp\left\{ |u|^{2} + |\upsilon|^{2} +
\text{Re}\, [\zeta^{\ast}(u^{2} + \upsilon^{2})] \right\} }{
\sqrt{1-|\zeta|^{2}} } \left[ e^{ 2\text{Re}\, y }
+ \tau^{2} e^{ -2\text{Re}\, y } + 2\tau
e^{-2|\upsilon|^{2}} \cos\left( \varphi -2\,\text{Im}\, y \right)
\right] ,   \label{5.32}
	\end{equation}
where
	\begin{equation}
y \equiv u^{\ast}\upsilon + \zeta^{\ast}u\upsilon =
\frac{ \upsilon z^{\ast} }{ \cosh s } .   \label{5.33}
	\end{equation}
All the properties of the displaced and squeezed superpositions
$|\upsilon,\tau,\varphi,\xi,z\rangle$ can be calculated by using
the analytic function $\Lambda(\upsilon,\tau,\varphi,\xi,z;\alpha)$
of Eq.\ (\ref{5.25}) or the Fock-state expansion of 
Eq.\ (\ref{5.29}).

\section{Squeezing and displacement of the SU(1,1) 
intelligent states}

Recently, Nieto and Truax \cite{NiTr} proposed a generalization
of squeezed states for an arbitrary dynamical symmetry group.
They found that the generalized squeezed states are eigenstates
of a linear combination of the lowering and raising generators
of a group. Actually, these states are the IS for the group 
Hermitian generators. Connections between the concepts of squeezing
and intelligence were further investigated by Trifonov 
\cite{Trif}. It turns out that the IS for two Hermitian generators
can provide an arbitrarily strong squeezing in either of these
observables \cite{Trif}. In the simplest case of the Heisenberg-Weyl
group $H_{3}$, the quadrature IS determined by the eigenvalue
equation (\ref{3.38}) are the canonical squeezed states 
$|\xi,\upsilon\rangle$ of Stoler and Yuen \cite{SS}.
By considering the $K_{1}$-$K_{2}$ IS, one can generalize the
concept of squeezing to the SU(1,1) group \cite{NiTr,PrAg,Trif}.
On the other hand, the usual squeezed vacuum states $|\xi\rangle$
are the generalized CS of SU(1,1). The algebra-eigenstate method
enables to treat both the generalized CS and the generalized 
squeezed states (i.e., the IS) for an arbitrary Lie group in a
unified way. Since the SU(1,1) Lie group in the two-photon 
realization (\ref{5.8}) is a subgroup of $H_{6}$, the SU(1,1) 
IS are a particular case of the two-photon AES. Furthermore, we can
consider the states generated by the squeezing transformations
$S(\xi)$ and displacement transformations $D(z)$ of the SU(1,1) IS.
Such states form a nonstandard set of the generalized two-photon CS.

According to Eq.\ (\ref{2.15}), the SU(1,1) IS are determined by
the eigenvalue equation 
	\begin{equation}
(\eta K_{1} - iK_{2}) |\lambda,\eta\rangle = 
\lambda |\lambda,\eta\rangle .   \label{6.1}
	\end{equation}
Here $\lambda$ is a complex eigenvalue and the parameter $\eta$ is
complex in the general case of the Robertson intelligence [an
equality is achieved in Eq.\ (\ref{2.13})] and real in the
particular case of the Heisenberg intelligence [an equality is 
achieved in Eq.\ (\ref{2.14})]. By evaluating the expectation values
over the state $|\lambda,\eta\rangle$, one gets \cite{Trif} (for
$\text{Re}\,\eta \neq 0$)
	\begin{equation}
\begin{array}{c}\vspace{0.2cm}
(\Delta K_{1})^{2} = \displaystyle{ \frac{\langle K_{0} \rangle}{
2\text{Re}\,\eta} } ,  \;\;\;\;\;\;\; 
(\Delta K_{2})^{2} = |\eta|^{2}  \displaystyle{ \frac{\langle K_{0} 
\rangle}{2\text{Re}\,\eta} } , \\
\sigma_{12} = \frac{1}{2} \langle K_{1}K_{2} + K_{2}K_{1} \rangle
- \langle K_{1} \rangle \langle K_{2} \rangle = \displaystyle{
\frac{\text{Im}\,\eta}{2\text{Re}\,\eta} } \langle K_{0} \rangle .
\end{array}    \label{6.2}
	\end{equation}
In the two-photon realization, the SU(1,1) Hermitian generators
$K_{1}$ and $K_{2}$ are given by Eq.\ (\ref{5.10}). Then the 
eigenvalue equation (\ref{6.1}) can be written in the form
	\begin{equation}
\left( \frac{\eta+1}{4} a^{2} + \frac{\eta-1}{4} a^{\dagger 2}
\right)  |\lambda,\eta\rangle = \lambda |\lambda,\eta\rangle .   
\label{6.3}
	\end{equation}
In the particular case $\eta=1$, the states $|\lambda,\eta\rangle$
are the eigenstates of the operator $a^{2}$, i.e., they reduce to
the coherent superpositions $|\upsilon,\tau,\varphi\rangle$
considered in the preceding section. Then, according to Eq.\ 
(\ref{6.2}), the uncertainties of $K_{1}$ and $K_{2}$ are equal
[cf. Eq.\ (\ref{5.13})]. In more general case of ordinary 
intelligent states ($\eta$ is real), the states 
$|\lambda,\eta\rangle$ are squeezed in $K_{1}$ for $\eta>1$ and
squeezed in $K_{2}$ for $\eta<1$. 

As usual, we define the entire analytic function 
	\begin{equation}
\Lambda(\lambda,\eta;\alpha) = e^{|\alpha|^{2}/2} \langle
\alpha^{\ast}|\lambda,\eta\rangle   \label{6.4}
	\end{equation}
that describes the IS $|\lambda,\eta\rangle$ in the Fock-Bargmann
representation. Then Eq.\ (\ref{6.3}) becomes a differential 
equation of the type (\ref{3.11}). The function 
$\Lambda(\lambda,\eta;\alpha)$ in this case is given by Eqs.\
(\ref{3.12}) and (\ref{3.16}) with the parameters
	\begin{equation}
\Delta^{2} = \frac{1}{4} (1-\eta^{2}) , \;\;\;\;\;\;\;
\sigma = 0 , \;\;\;\;\;\;\; \mu_{\Delta} = 0 , \;\;\;\;\;\;\;
d = \displaystyle{ \frac{1}{4} - \frac{\lambda}{2\Delta} } .
\label{6.5}
	\end{equation}
Therefore we obtain
\begin{mathletters}  \label{6.6}
	\begin{eqnarray}
& & \Lambda_{e}(\lambda,\eta;\alpha) = {\cal A}(\alpha)\,
{}_{1}\! F_{1} \left( \frac{1}{4} 
- \frac{\lambda}{2\Delta} \left| \frac{1}{2} \right| - \Omega_{\eta} 
\alpha^{2}  \right) , \label{6.6a} \\
& & \Lambda_{o}(\lambda,\eta;\alpha) = \alpha {\cal A}(\alpha)\,
{}_{1}\! F_{1} \left( \frac{3}{4} 
- \frac{\lambda}{2\Delta} \left| \frac{3}{2} \right| - \Omega_{\eta} 
\alpha^{2}  \right) ,  \label{6.6b}
	\end{eqnarray}
\end{mathletters}
where
	\begin{eqnarray}
& & {\cal A}(\alpha) \equiv \exp\left( \mbox{$\frac{1}{2}$}
\Omega_{\eta}\alpha^{2} \right) ,   \label{6.7A}  \\
& & \Omega_{\eta}^{2} \equiv \frac{\Delta^{2}}{4\beta_{2}^{2}}
= \frac{1-\eta}{1+\eta} .  \label{6.7}
	\end{eqnarray}
The solutions $\Lambda_{e}$ and $\Lambda_{o}$ represent the states 
belonging to the SU(1,1) irreducible sectors spanned by the Fock
states $|n\rangle$ with even and odd values of $n$, respectively.
The total solution is given by a superposition of $\Lambda_{e}$ and 
$\Lambda_{o}$. Note that the double-valuedness of $\Delta$ and
$\Omega_{\eta}$ reflects the invariance of the solution under 
Kummer's transformation (\ref{4.15}). The normalization condition
(\ref{3.6}) requires $|\Omega_{\eta}|<1$ which is satisfied for
$\text{Re}\,\eta>0$. This is the only restriction on values of
$\eta$. If we express the Kummer functions ${}_{1}\! F_{1}\left(
d\left|\frac{1}{2}\right|x\right)$ and ${}_{1}\! F_{1}\left(
d+\frac{1}{2}\left|\frac{3}{2}\right|x\right)$ in terms of the
parabolic cylinder functions $D_{-2d}(\pm x)$ by means of Eq.\
(\ref{pcf}), we will recover the results of Prakash and Agarwal
\cite{PrAg}.

An important feature of the algebra-eigenstate method is the 
possibility to find relations between various types of states.
For the SU(1,1) group, the standard set of the generalized CS have 
an intersection with the set of the ordinary IS \cite{WodEb,BBA:jpa}
and is a subset of the wider set of the generalized IS 
\cite{Trif}. Let us demonstrate these relations by using the
Fock-Bargmann representation of the two-photon AES. The squeezed
vacuum states $|\xi\rangle$ which are the standard CS of SU(1,1)
are represented by the function $\Lambda(\xi,\upsilon;\alpha)$
of Eq. (\ref{3.37}) with $\upsilon=0$, i.e., 
	\begin{equation}
\Lambda(\xi;\alpha) = (1-|\zeta|^{2})^{1/4} \exp\left(
\mbox{$\frac{1}{2}$} \zeta\alpha^{2} \right) .  \label{6.8}
	\end{equation}
On the other hand, when
	\begin{equation}
\frac{1}{4} - \frac{\lambda}{2\Delta} = \frac{1}{2} , \label{6.9}
	\end{equation}
the formula ${}_{1}\! F_{1}(d|d|x)=e^{x}$ enables us to write 
Eq.\ (\ref{6.6a}) in the (normalized) form  
	\begin{equation}
\Lambda_{e}(\lambda,\eta;\alpha) = (1-|\Omega_{\eta}|^{2})^{1/4} 
\exp\left( -\mbox{$\frac{1}{2}$} \Omega_{\eta}\alpha^{2} \right) . 
\label{6.10}
	\end{equation}
Therefore, the intelligent state $|\lambda,\eta\rangle$ is the
standard coherent state $|\xi\rangle$ under the condition
	\begin{equation}
\lambda = -\Delta/2 = \pm \frac{1}{4} \sqrt{1-\eta^{2}} .
\label{6.11}
	\end{equation}
The corresponding coherent-state amplitude is
	\begin{equation}
\zeta = -\Omega_{\eta} = \pm \sqrt{ \frac{1-\eta}{1+\eta} } .
\label{6.12}
	\end{equation}
The condition $|\zeta|<1$ is guaranteed by virtue of the 
normalization requirement $|\Omega_{\eta}|<1$ ($\text{Re}\,\eta>0$).
When $\eta$ is complex (the case of the generalized IS), $\zeta$
can acquire any value in the unit disk. It means that the standard
CS form a subset of the generalized IS. However, when $\eta$ is
real (the case of the ordinary IS), $\zeta$ is real for $\eta<1$
and pure imaginary for $\eta>1$. It means that the standard set
of the generalized CS has an intersection with the set of the
ordinary IS. The standard CS are the ordinary IS squeezed in $K_{2}$
for real $\zeta$ and squeezed in $K_{1}$ for pure imaginary $\zeta$.

Now, let us consider the action of the squeezing operator $S(\xi
= se^{i\theta})$. By applying $S(\xi)$ to Eq.\ (\ref{6.3}), we 
obtain the eigenvalue equation	
	\begin{equation}
\left( \frac{\eta+1}{4} a_{\xi}^{2} + \frac{\eta-1}{4} 
a_{\xi}^{\dagger 2} \right)  |\lambda,\eta,\xi\rangle = 
\lambda |\lambda,\eta,\xi\rangle     \label{6.13}
	\end{equation}
satisfied by the squeezed IS
	\begin{equation}
|\lambda,\eta,\xi\rangle = S(\xi) |\lambda,\eta\rangle .
\label{6.14}
	\end{equation}
The operator $a_{\xi}$ is given by Eq.\ (\ref{5.16}) and 
$a_{\xi}^{\dagger}$ is its Hermitian conjugate. Equation 
(\ref{6.13}) can be written in the standard form: 
	\begin{equation}
(\beta_{1}N + \beta_{2}a^{2} + \beta_{3}a^{\dagger 2})
|\lambda,\eta,\xi\rangle = \lambda_{\xi} |\lambda,\eta,\xi\rangle ,
\label{6.15}
	\end{equation}
where
	\begin{equation}
\begin{array}{c}\vspace{0.2cm}
\beta_{1} = -2\zeta(\eta+1)-2\zeta^{\ast}(\eta-1) , 
\;\;\;\;\;\;\;
\beta_{2} = (\eta+1)+\zeta^{\ast 2}(\eta-1) , \\ \vspace{0.2cm}
\beta_{3} = \zeta^{2}(\eta+1)+(\eta-1) ,
\end{array}   \label{6.16}
	\end{equation}\vspace*{-1.0cm}
	\begin{equation}
\lambda_{\xi} = 4(1-|\zeta|^{2})\lambda + \zeta(\eta+1)
+ \zeta^{\ast}(\eta-1) ,   \label{6.17}
	\end{equation}
and $\zeta=\tanh s\, e^{i\theta}$ is defined by Eq.\ (\ref{3.zeta}).
The analytic function $\Lambda(\lambda,\eta,\xi;\alpha)$ 
representing the squeezed IS is given by Eqs.\ (\ref{3.12}) and
(\ref{3.16}) with the parameters
	\begin{equation}
\Delta^{2} = 4(1-\eta^{2})(1-|\zeta|^{2}) , \;\;\;\;\;\;\;
\sigma = 0 , \;\;\;\;\;\;\; \mu_{\Delta} = 0 , \;\;\;\;\;\;\;
d = \frac{1}{4} - 2(1-|\zeta|^{2})\lambda/\Delta  .
\label{6.18}
	\end{equation}
The unitary squeezing operator $S(\xi)$ is an element of the
SU(1,1) group and, therefore, Eq.\ (\ref{6.15}) does not include
the first-order operators $a$ and $a^{\dagger}$ which represent
one-photon processes. Then the total solution 
$\Lambda(\lambda,\eta,\xi;\alpha)$ can be written as a 
superposition of two solutions $\Lambda_{e}$ and $\Lambda_{o}$
which represent the two irreducible sectors of SU(1,1).

At the next step, we apply the displacement operator $D(z)$. The
resulting eigenvalue equation is 
	\begin{equation}
\left( \frac{\eta+1}{4} a_{\xi,z}^{2} + \frac{\eta-1}{4} 
a_{\xi,z}^{\dagger 2} \right)  |\lambda,\eta,\xi,z\rangle = 
\lambda |\lambda,\eta,\xi,z\rangle ,    \label{6.19}
	\end{equation}
where
	\begin{equation}
|\lambda,\eta,\xi,z\rangle = D(z) S(\xi) |\lambda,\eta\rangle .
\label{6.20}
	\end{equation}
are the displaced and squeezed IS.
The operator $a_{\xi,z}$ is given by Eq.\ (\ref{5.20}) and 
$a_{\xi,z}^{\dagger}$ is its Hermitian conjugate. Equation 
(\ref{6.19}) can be written in the standard form:
	\begin{equation}
(\beta_{1}N + \beta_{2}a^{2} + \beta_{3}a^{\dagger 2} + \beta_{4}a 
+ \beta_{5}a^{\dagger}) |\lambda,\eta,\xi,z\rangle 
= \lambda_{\xi,z} |\lambda,\eta,\xi,z\rangle  ,
\label{6.21}
	\end{equation}
where $\beta_{1}$, $\beta_{2}$ and $\beta_{3}$ remain as given by
Eq.\ (\ref{6.16}), and 
	\begin{equation}
\beta_{4} = -2\rho(\eta+1)+2\zeta^{\ast}\rho^{\ast}(\eta-1) , 
\;\;\;\;\;\;\;\;
\beta_{5} = 2\zeta\rho(\eta+1)-2\rho^{\ast}(\eta-1) , 
\label{6.22}
	\end{equation}\vspace*{-1.2cm}
	\begin{equation}
\lambda_{\xi,z} = 4(1-|\zeta|^{2})\lambda + 
(\zeta-\rho^{2})(\eta+1) 
+ (\zeta^{\ast}-\rho^{\ast 2})(\eta-1) .   \label{6.23}
	\end{equation}
Here $\rho = z-\zeta z^{\ast}$ is defined by Eq.\ (\ref{5.22}).
The analytic function $\Lambda(\lambda,\eta,\xi,z;\alpha)$ 
representing the displaced and squeezed IS is given by general 
equations (\ref{3.12}) and (\ref{3.16}). The corresponding
parameters are
	\begin{equation}
\Delta^{2} = 4(1-\eta^{2})(1-|\zeta|^{2}) , \label{6.24}
	\end{equation}\vspace*{-1.0cm}
	\begin{equation}
\sigma = \frac{ \frac{1}{2}z^{\ast}(1-|\zeta|^{2})\Delta^{2}
+[-\rho(\eta+1)+\zeta^{\ast}\rho^{\ast}(\eta-1)]\Delta }{
(\eta+1) + \zeta^{\ast 2}(\eta-1) } ,   \label{6.25}
	\end{equation}\vspace*{-1.0cm}
	\begin{equation}
\mu_{\Delta} = \rho^{\ast}+\zeta^{\ast}\rho = 
z^{\ast}(1-|\zeta|^{2}) , \label{6.26}
	\end{equation}
and $d$ can be found from the general expression (\ref{3.18}).

The analytic function $\Lambda(\lambda,\eta,z;\alpha)$ that
represents the displaced IS
	\begin{equation}
|\lambda,\eta,z\rangle = D(z) |\lambda,\eta\rangle 
\label{6.27}
	\end{equation}
can be found by taking $\zeta=0$ in the expressions for the 
displaced and squeezed IS. Then we obtain
	\begin{equation}
\begin{array}{c} 
\beta_{1} = 0 , \;\;\;\;\;\;\;
\beta_{2} = (\eta+1) , \;\;\;\;\;\;\;
\beta_{3} = (\eta-1) , \\
\beta_{4} = -2z(\eta+1), \;\;\;\;\;\;\;
\beta_{5} = -2z^{\ast}(\eta-1) , 
\end{array}   \label{6.28}
	\end{equation}\vspace*{-0.8cm}
	\begin{equation}
\lambda_{z} = 4\lambda - z^{2}(\eta+1)-z^{\ast 2}(\eta-1) .   
\label{6.29}
	\end{equation}
By using these results, we find the usual set of the parameters:
	\begin{equation}
\begin{array}{c}
\Delta^{2} = 4(1-\eta^{2}) , \;\;\;\;\;\;\;
\sigma = 2z^{\ast}(1-\eta) - z\Delta ,  \;\;\;\;\;\;\;
\mu_{\Delta} = z^{\ast} , \\
d = \frac{1}{4} + [z^{2}(\eta+1)-2\lambda]/\Delta  .
\end{array}    \label{6.30}
	\end{equation}

Let us finish this discussion by a brief review of possibilities
for the generation of the SU(1,1) IS. Gerry and Hach \cite{GeHa} 
demonstrated a possibility to generate coherent superposition 
states for the long-time evolution of the competition between
two-photon absorption and two-photon parametric processes for
a special initial state. This method can be also applied to the
production of the SU(1,1) IS more general than the coherent 
superposition states. Prakash and Agarwal \cite{PrAg} proposed
to use the degenerate down-conversion of coherent light in presence
of a broadband squeezed field in the cavity (see also Ref.
\cite{GeGr} where the same idea was applied to the generation
of the two-mode SU(1,1) IS). This method is based on an earlier
proposal \cite{AgPu} introduced in the context of the SU(2) group.

\section{Conclusions}

We have shown that almost all photon states known in the context
of the two-photon algebra can be considered as the AES. Therefore,
the algebra-eigenstate formalism unifies the description of various
types of states within a common frame. This helps in understanding
of relations between different kinds of states and of the physical
basis of their mathematical properties. The theory of the AES is
in general applicable to an arbitrary Lie group and will be useful
for a unified description of generalized coherence and squeezing
in a wide class of quantum systems. In the present work we have
concentrated on the basic quantum optical phenomena, such as usual
displacement and squeezing of the quantized single-mode light field.
The mathematical formulation of these physical processes is based
on the two-photon group $H_{6}$. The corresponding two-photon 
AES form an extremely wide set that includes as particular cases 
various types of photon states whose properties have drawn recently 
considerable attention. The standard CS of Glauber, two-photon CS
(canonical squeezed states) of Stoler and Yuen, displaced and 
squeezed Fock states, displaced and squeezed coherent 
superpositions, and displaced and squeezed SU(1,1) IS are 
incorporated into the set of the two-photon AES. The Fock-Bargmann
analytic representation of all the particular subsets is obtained
from the general differential equation (\ref{3.11}) that is common 
for all the kinds of the AES. Then the powerful theory of analytic 
functions can be used for investigating properties of various types 
of states and relations between them.

\acknowledgments

The author thanks Profs. A. Mann, M. S. Marinov, and Y. Ben-Aryeh
for interesting and stimulating discussions. The financial help
from the Technion is gratefully acknowledged.

\end{document}